\documentclass[prl,twocolumn,reprint,preprintnumbers,nofootinbib,superscriptaddress]{revtex4-2}
\input{header.tex}

\begin{document}

\reportnum{-2}{CERN-TH-2024-188}

\title{New physics decaying into metastable particles: impact on cosmic neutrinos}

\author{Kensuke~Akita}
\email{kensuke@hep-th.phys.s.u-tokyo.ac.jp}
\affiliation{Department of Physics, The University of Tokyo, Bunkyo-ku, Tokyo 113-0033, Japan}
\author{Gideon~Baur}
\email{gbaur@uni-bonn.de}
\affiliation{Institut für Astroteilchen Physik, Karlsruher Institut für Technologie (KIT), Hermann-von-Helmholtz-Platz 1, 76344 Eggenstein-Leopoldshafen, Germany}
\affiliation{Bethe Center for Theoretical Physics, Universität Bonn, D-53115, Germany}
\author{Maksym~Ovchynnikov}
\email{maksym.ovchynnikov@cern.ch}
\affiliation{Theoretical Physics Department, CERN, 1211 Geneva 23, Switzerland}
\affiliation{Institut für Astroteilchen Physik, Karlsruher Institut für Technologie (KIT), Hermann-von-Helmholtz-Platz 1, 76344 Eggenstein-Leopoldshafen, Germany}
\author{Thomas Schwetz}
\email{schwetz@kit.edu}
\affiliation{Institut für Astroteilchen Physik, Karlsruher Institut für Technologie (KIT), Hermann-von-Helmholtz-Platz 1, 76344 Eggenstein-Leopoldshafen, Germany}
\author{Vsevolod Syvolap}
\email{sivolapseva@gmail.com}
\affiliation{Instituut-Lorentz, Leiden University, Niels Bohrweg 2, 2333 CA Leiden, The Netherlands}
\date{\today}

\begin{abstract}
We investigate decays of hypothetical unstable new physics particles into metastable species such as muons, pions, or kaons in the Early Universe, when temperatures are in the MeV range, and study how they affect cosmic neutrinos. We demonstrate that the non-trivial dynamics of metastables in the plasma alters the impact of the new physics particles on the neutrino population, including the effective number of neutrino degrees of freedom, $N_{\rm eff}$, modifies neutrino spectral distortions, and may induce asymmetries in neutrino and antineutrino energy distributions. These modifications have important implications for observables such as Big Bang Nucleosynthesis and the Cosmic Microwave Background, especially in light of upcoming CMB observations aiming to reach percent-level precision on $N_{\rm eff}$. We illustrate our findings with a few examples of new physics particles and provide a computational tool available for further exploration.
\end{abstract}

\maketitle

\textbf{Introduction.} The thermal plasma of the Early Universe is a sensitive probe of new physics. In particular, any modifications of the standard evolution in the period when neutrinos decouple from the thermal bath at temperatures $T \lesssim 5\text{ MeV}$ can alter primordial neutrino properties~\cite{Dolgov:2002wy}, which then may affect key cosmological observables, including primordial nuclear abundances~\cite{Sarkar:1995dd,Dolgov:2000jw,Kohri:2001jx,Hannestad:2004px,Pospelov:2010hj,Kawasaki:2017bqm,Boyarsky:2020dzc}, Cosmic Microwave Background (CMB)~\cite{Ellis:1984eq,Moroi:1993mb,Kawasaki:1994af,Giudice:2000ex,Hannestad:2004px,Kanzaki:2007pd,Fradette:2017sdd,Fradette:2018hhl,Hasegawa:2019jsa,Sabti:2020yrt,Boyarsky:2021yoh,Mastrototaro:2021wzl,Rasmussen:2021kbf}, and constraints on neutrino masses~\cite{Alvey:2021sji,Alvey:2021xmq,Escudero:2022gez,Naredo-Tuero:2024sgf}.

A common scenario with new physics involves beyond the Standard Model Long-Lived Particles (LLPs) with lifetimes $\tau_{X} \lesssim 1\,\text{s}$ decaying into \emph{metastable} Standard Model (SM) particles ($Y = \mu^\pm,\pi^\pm,K^\pm,K_L$)~\cite{Beacham:2019nyx,Boiarska:2019jym,Bauer:2020jbp,Ilten:2018crw,Bondarenko:2018ptm}. When these $Y$ particles subsequently decay themselves, they inject high-energy neutrinos, which cause two independent effects. First, they affect the effective number of relativistic neutrino species, $N_{\rm eff}$, defined as
\begin{equation}
N_{\rm eff} = \frac{8}{7}\left(\frac{11}{4}\right)^{\frac{4}{3}}\frac{\rho_{\text{UR}}-\rho_{\gamma}}{\rho_{\gamma}}\bigg|_{m_{\nu}\ll T \ll m_{e}},
\end{equation}
where $\rho_{\text{UR}},\rho_{\gamma}$ are the energy densities of all ultra-relativistic particles and photons, respectively, $T$ is the electromagnetic (EM) plasma temperature, and $m_{\nu}, m_{e}$ are the masses of neutrinos and electrons. Second, they induce spectral distortions~\cite{Boyarsky:2021yoh,Rasmussen:2021kbf,Ovchynnikov:2024rfu,Ovchynnikov:2024xyd}. The latter is important for the proton-to-neutron conversion, which defines the onset of Big Bang Nucleosynthesis. Also, they break the degeneracy between $N_{\rm eff}$ and the number density of neutrinos, affecting the role of neutrino mass in cosmology after they become non-relativistic. 

Previous studies~\cite{Fradette:2017sdd,Fradette:2018hhl,Gelmini:2020ekg,Sabti:2020yrt,Boyarsky:2021yoh,Rasmussen:2021kbf} analyzing the impact of LLPs on cosmic neutrinos have assumed that the metastable particles always decay after thermalizing. In this letter, we demonstrate that if injected at MeV temperatures, $Y$\!s can disappear by efficiently annihilating or interacting with nucleons before decaying, which qualitatively changes the evolution of neutrinos. Hence, this discovery has significant implications for constraining or discovering new physics through cosmological observations. Our approach applies to a wide range of new physics scenarios, including vanilla decaying LLPs, low-temperature reheating scenarios with hadronically decaying particles, and low-temperature baryogenesis models~\cite{Aitken:2017wie,Elor:2018twp}.

Detailed methodologies, comprehensive analyses, extended case studies, and the description of the code handling the dynamics of metastables\footnote{Available on \faGithub \cite{GitHub-metastable} and Zenodo~\cite{Zenodo}.} and the solver of neutrino Boltzmann equation incorporating this dynamics\footnote{Available on \faGithub \cite{GitHub-solver}.} are provided in the Companion Paper~\cite{Akita:2024ork}.

\begin{figure*}[t!]
    \centering
    \includegraphics[width=0.9\textwidth]{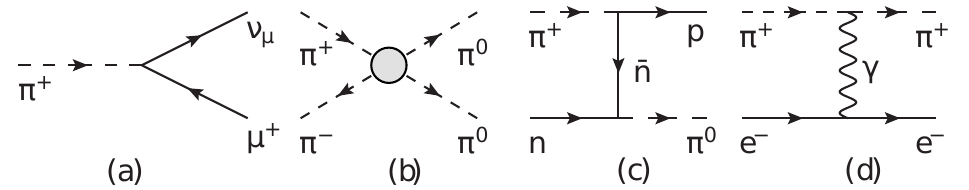}
    \caption{Interaction processes of the injected metastable particles $Y$ in the MeV primordial plasma: decay (a), annihilation with the injected antiparticle (b), interaction with nucleons (c), and elastic EM scattering (d). The process (a) injects non-thermal neutrinos, the reactions (b), (c) lead to the disappearance of $Y$\!s without decaying, injecting energy either to the electromagnetic plasma or to lighter $Y$\!s, whereas (d) places most of the $Y$\!s kinetic energy in the EM plasma.}
    \label{fig:interactions}
\end{figure*}

\bigskip

\textbf{Dynamics of metastable particles.} The rates of several processes involving metastable particles are significantly larger than the Hubble expansion rate at MeV temperatures, which leads to a complicated evolution in the primordial plasma. The processes include (see fig.~\ref{fig:interactions}):

\begin{itemize}
    \item[(a)] \textit{Decay}: $Y \to \text{SM particles}$. Decays are governed by weak interactions. As a result, the lifetimes $\tau_Y \sim 10^{-8} - 10^{-6}\text{ s}$ are not short enough to neglect the possibility of various scattering processes with $Y$\!s prior to the decay. Decay products of $Y$\!s include high-energy non-thermal neutrinos, which lead to neutrino spectral distortions.
    \item[(b)] \textit{Annihilation}: $Y + \bar{Y} \to \text{SM particles}$. Both particles and antiparticles participating in the process originate from decays of LLPs. The process is driven by electromagnetic or strong forces, and the largeness of the cross-section compensates for the smallness of the $\bar{Y}$ yield available for the annihilation with $Y$.
    \item[(c)] \textit{Interaction with nucleons $\mathcal{N}$}: various quasi-elastic processes of the type $Y + \mathcal{N} \to \mathcal{N}^{(\prime)} + \text{other particles}$. Examples are $\pi^- + p \to n + \pi^0$ and $K^- + p \to n + 2\pi$, changing the nucleon type, and $K^{-}+p\to p + 2\pi$, that leave it unchanged. The process's rate is parametrically suppressed by the nucleon number density. Because of this, this interaction is only efficient in the case of mesonic $Y$\!s, as then the smallness of the baryon number is compensated by a large interaction cross-section driven by the strong force.\footnote{The $p\leftrightarrow n$ processes have been included in the works~\cite{Reno:1987qw,Kohri:2001jx,Kawasaki:2004qu,Pospelov:2010cw,Kawasaki:2017bqm,Hasegawa:2019jsa,Boyarsky:2020dzc}, studying the impact of various scenarios with LLPs decaying into $Y$\!s on primordial nuclear abundances. However, to the best of our knowledge, they have not been included in any previous study of the impact on neutrinos.}
    \item[(d)] \textit{Elastic electromagnetic scatterings}: $Y+\text{EM}\to Y+\text{EM}$. It transfers the kinetic energy of the charged $Y$ particles to the EM plasma, leading to the thermalization of the kinetic energy of $Y$s with photons and electrons. This process is typically the most efficient one, as both the number density of interacting counterparts and the cross-section are large.
\end{itemize}
Processes (b)-(d) do not directly inject energy into the neutrino sector. Consequently, when annihilation and interactions with nucleons dominate over decays, the metastable particles transfer all their energy to the electromagnetic sector and lighter metastable particles (which, similarly, disappear before decaying and release the energy to the EM plasma) instead of producing high-energy neutrinos.

To quantify the impact of these processes on primordial neutrinos, we have implemented a two-step analysis and presented it in the publicly available codes. First, we have solved the coupled Boltzmann equations governing the number densities of $Y$ particles and nucleons in the presence of the decaying LLPs. Second, we have incorporated the resulting dynamics into the source term of the collision integral for the solver of the unintegrated (i.e., momentum-dependent) Boltzmann equations on the neutrino distribution functions (also including neutrino oscillations). For a detailed description of the methodology and cross-section calculations, see our Companion Paper~\cite{Akita:2024ork}.

\bigskip

\textbf{Impact on the properties of neutrinos.} The suppression of $Y$ decays due to annihilation and interactions with nucleons alters the expected neutrino properties. We summarize them below:

\begin{itemize}
    \item \textit{Effective number of relativistic neutrino species ($N_{\text{\rm eff}}$):} As there is less energy injection into the neutrino sector, there is a decrease in $N_{\rm eff}$ compared to the setup where $Y$ decays are inevitable.
    
    \item \textit{Neutrino spectral distortions:} Less $Y$ decays imply fewer injected high-energy neutrinos, and hence, there is no enhanced high-energy neutrino tail, implying smaller spectral distortions.
    
    \item \textit{Neutrino-antineutrino energy distribution asymmetry:}
    The dynamics of $K^{+}$ and $K^{-}$ are not symmetric: whereas $K^{-}$ may efficiently disappear because of the interactions with nucleons, there is no such a process for $K^{+}$.\footnote{This is because the thresholdless scatterings occur via intermediate resonances $\Lambda, \Sigma$. The processes involving $K^{+}$ require the resonances with positive baryon number and strangeness that do not exist~\cite{Reno:1987qw}.} As a result, $K^{+}$ decays more often than $K^{-}$. It leads to producing more high-energy neutrinos than antineutrinos in the energy range $E_{\nu}> m_{\mu}/2$. On the other hand, the same reason leads to an excess of $\mu^{+},\pi^{+}$, which induces more antineutrinos than neutrinos in the energy range $E_{\nu}<m_{\mu}/2$. Hence, although a neutrino-antineutrino asymmetry in number densities is bounded due to lepton, baryon, and electric charge conservation, the resulting asymmetry in their energy distributions may be sizeable, as the interactions with nucleons are very efficient.
\end{itemize}

These modifications may have profound implications for Big Bang Nucleosynthesis (BBN) and CMB anisotropies. Particularly, the shape of the neutrino distribution, as well as a possible neutrino/antineutrino asymmetry, is important for the proton-to-neutron conversion rates (determining the onset of BBN) and the energy density of non-relativistic neutrinos.

\bigskip

\textbf{Case studies.} To demonstrate the impact of the $Y$ dynamics on the neutrino properties, we consider two models with LLPs $X$: a toy model where $X$ decays solely into pions and Higgs-like scalars.

\begin{figure}[t!]
    \centering
    \includegraphics[width=0.45\textwidth]{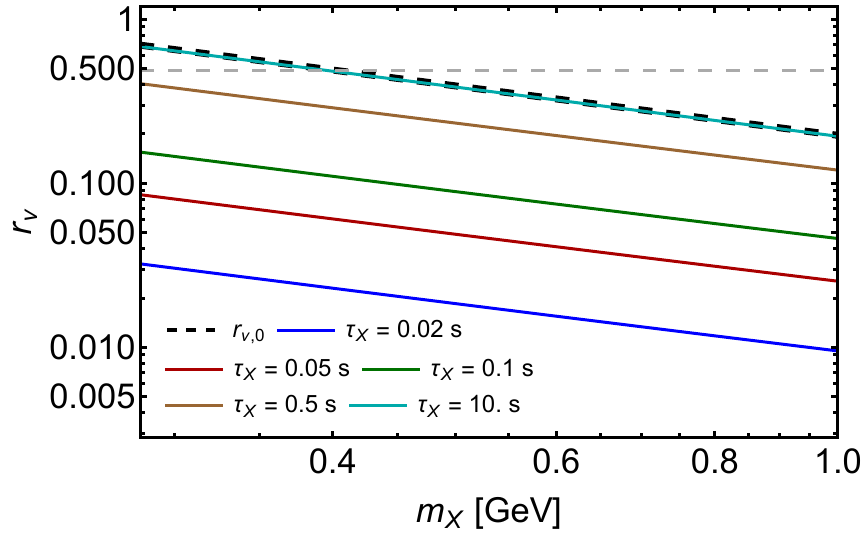} \\ 
    \includegraphics[width=0.45\textwidth]{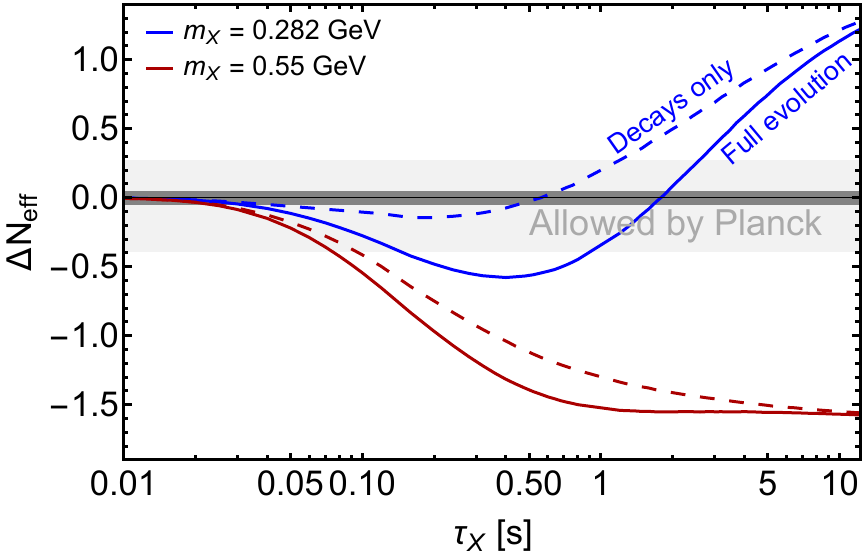}
    \caption{Impact of the particle $X$ decaying solely into charged pions as a function of its mass $m_{X}$ and lifetime $\tau_{X}$ on neutrinos. The plots are obtained considering the non-trivial evolution of metastable particles $Y$. \textit{Top panel}: qualitative impact of disappearance of pions and muons before decaying -- cumulative fraction of the energy from $X$ decays directly injected into neutrinos, $r_{\nu}$, as a function of the $X$ mass for different $X$ lifetimes. The black dashed line ($r_{\nu,0}$) corresponds to the case when all $Y$\!s particles inevitably decay, whereas solid lines take into account annihilations and nucleon interactions of $Y$s. \textit{Bottom panel}: correction to the effective number of neutrino species, $\Delta N_{\rm eff} \equiv N_{\rm eff}-N_{\rm eff}^{\Lambda\text{CDM}}$ for two representative masses. The results are obtained using the unintegrated neutrino Boltzmann equations solver with the incorporated dynamics of the metastable particles. The solid lines include \textit{both} annihilations and decays of $Y$ particles, whereas the dashed curves assume \textit{only} inevitable decays. The gray band represents the Planck 95\%CL constraints $N_{\rm eff} = 2.99^{+0.33}_{-0.34}$~\cite{Planck:2018vyg}, whereas the black band shows the forecasted sensitivity of the Simons Observatory, which we assume to be centered at $\Delta N_{\rm eff}$ = 0~\cite{SimonsObservatory:2018koc}.}
    \label{fig:DeltaNeff-toy}
\end{figure}

\textit{Toy model.} In this case, we fix the LLP abundance and branching ratios, allowing the mass $m_X$ and lifetime $\tau_X$ to vary. The abundance is chosen as
\begin{equation}
    \mathcal{Y}_{X} \equiv \left( \frac{n_{\text{LLP}}}{s}\right)_{T = 10\text{ MeV}} = 2\cdot 10^{-3},
\end{equation}
which corresponds to a scenario where the LLP was in thermal equilibrium and decoupled while still relativistic. Here we consider the range of masses $2\cdot m_{\pi} < m_{X} < 1 \text{ GeV}$ and lifetimes $10^{-2} \text{ s} < \tau_{X} < 10 \text{ s}$. The only decay mode of $X$ is into a pair of charged pions. However, they can subsequently decay into muons, i.e., they have a coupled evolution.

Fig.~\ref{fig:DeltaNeff-toy} illustrates the impact of the LLP decays on neutrinos and how the non-trivial dynamics of metastable particles affects it, considering various LLP mass and lifetime. In the top panel, we show the ratio of the energy injected directly into neutrinos to the total energy released by decaying LLP:
\begin{align}
    r_{\nu} &= \left. \frac{\rho_{\text{inj}, \nu}}{\rho_{\text{inj}}} \right|_{t=\infty}.
    \label{eq:rnu} 
\end{align}
To concentrate on the metastables dynamics, it does not incorporate interactions between neutrinos and electromagnetic particles. In the absence of the direct LLP decays into neutrinos, $r_{\nu}$ is determined by the decays of pions and muons. It reaches the maximal possible value, $r_{\nu,0}$, when decays are inevitable. However, at MeV temperatures, $Y$\!s instead prefer to disappear via annihilations and interactions with nucleons. It leads to a drop $r_\nu < r_{\nu,0}$. In particular, for LLP lifetimes $\tau_X \ll 1$~s, it may result in a tiny $r_{\nu}=\mathcal{O}(1\%)$. As the lifetime increases, annihilation and scattering off nucleons become inefficient, more and more $Y$\!s decay, and for $\tau_X \gtrsim 1$~s $r_{\nu}$ approaches $r_{\nu,0}$. 

This qualitative behavior allows an understanding of the bottom panel of the figure. It shows the deviation of the effective neutrino species from its standard value: $\Delta N_{\rm eff} \equiv N_{\rm eff} - N_{\rm eff}^{\Lambda\text{CDM}}$, which is calculated using the complete neutrino Boltzmann solver incorporating the dynamics of $Y$\!s. We show $\Delta N_{\rm eff}(\tau_{X})$ for two representative values of LLP masses, chosen such that the corresponding limiting values $\lim_{\tau_{X}\gg 1\text{ s}}\Delta N_{\rm eff}$ are opposite: a positive and a negative. 

To highlight the importance of the dynamics, we also consider two setups: when all metastables inevitably decay, and when full dynamics is included. In both cases, $\Delta N_{\rm eff}$ is negative in the domain $\tau_{X}\lesssim 1\text{ s}$; this is explained either by the non-trivial thermalization of high-energy neutrinos~\cite{Boyarsky:2021yoh,Ovchynnikov:2024rfu,Ovchynnikov:2024xyd,Akita:2024ork} (the first setup) or the disappearance of metastable particles before their decays (the second setup). However, for the second setup, the value of $\Delta N_{\rm eff}$ becomes significantly lower, in some regions even changing its sign compared to the setup when only decays are taken into account. We compare the size of the effect with the present accuracy on $\Delta N_{\rm eff}$ measurements from Planck, as well as the sensitivity of the future Simons Observatory. It clearly shows that the impact of the effect pointed out here is comparable to present uncertainties and much larger than future sensitivities. 

\begin{figure}[t!]
    \centering
    \includegraphics[width=\linewidth]{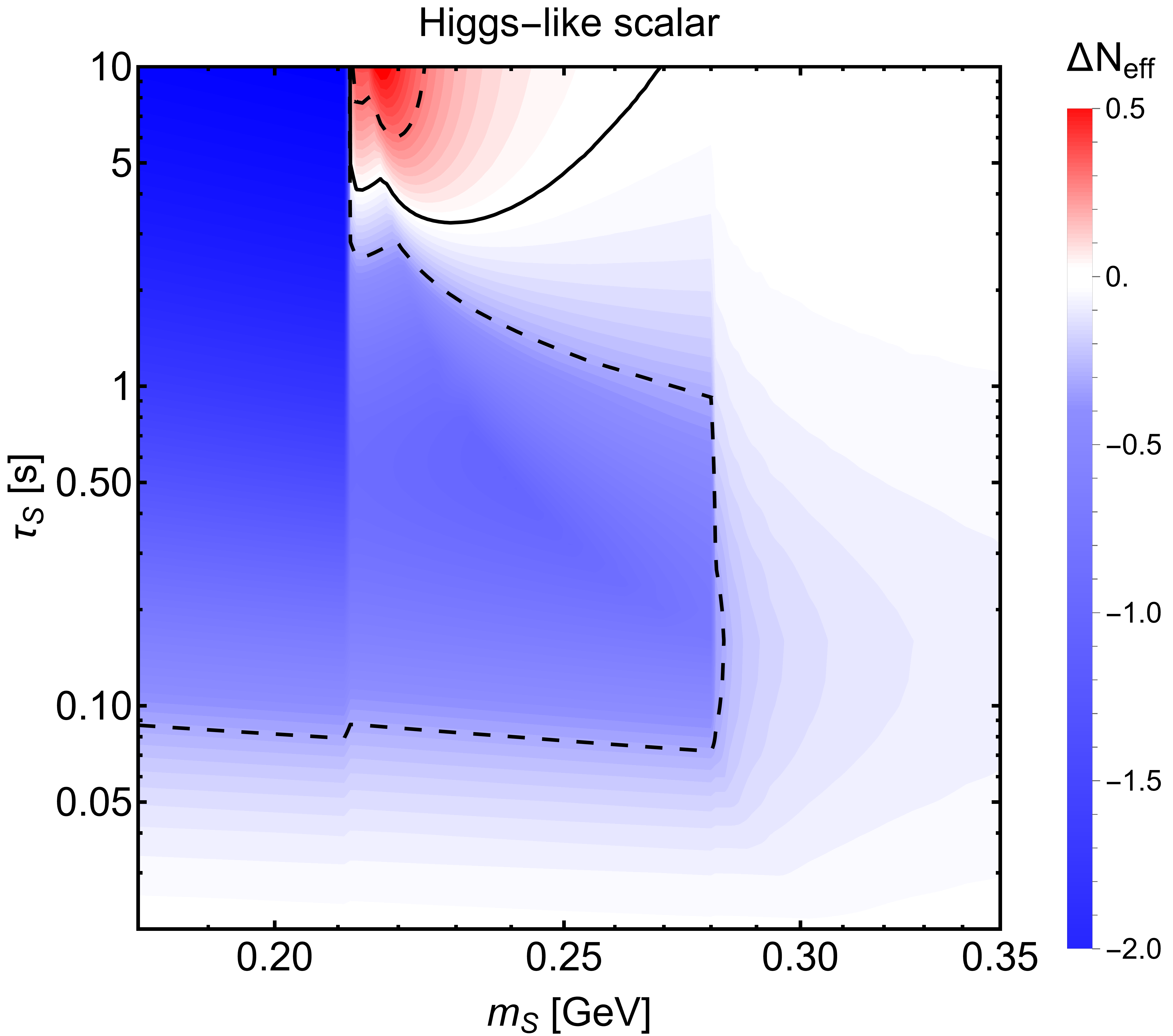}
    \caption{Effect of Higgs-like scalars on $\Delta N_{\rm eff}$ in the parameter space of the scalar mass $m_{S}$ and lifetime $\tau_{S}$. The solid black line marks the parameter space where $\Delta N_{\rm eff} = 0$, while dashed lines indicate regions where $\Delta N_{\rm eff}$ exceeds the Planck 95\% CL bound~\cite{Planck:2018nkj}. Similarly to the bottom panel of fig.~\ref{fig:DeltaNeff-toy}, the plots have been obtained using the unintegrated neutrino Boltzmann solver. In the mass range $2m_{\mu}<m_{S}<2m_{\mu}$, $\Delta N_{\rm eff}$ experiences the transition between negative and positive values (as in fig.~\ref{fig:DeltaNeff-toy}).}
    \label{fig:Neff-scalar}
\end{figure}

For the toy model study, we arbitrarily imposed two simplifications to make the analysis transparent. First, we assumed that its abundance is independent of mass and lifetime. This may not be the case for particular LLPs. If one interaction constant handles both the production and decays of LLPs, its abundance is uniquely fixed by specifying $m_X$ and $\tau_X$. Second, depending on the LLP, it may have multiple decay modes, including EM particles, the $Y$\!s, and neutrinos.

This is the case for our following example -- the \emph{Higgs-like scalars}, denoted by $S$ (more models are considered in our Companion Paper~\cite{Akita:2024ork}). They behave as a light Higgs boson with the couplings suppressed by the mixing angle $\theta$. If this model is realized in nature, the $S$\!s have been produced in the Early Universe at temperatures $T=\mathcal{O}(100\text{ GeV})$. In the GeV mass range, their number density-to-entropy ratio before decaying is $Y \approx 5\cdot 10^{11}\theta^{2}$~\cite{Fradette:2018hhl}. 

The scalars predominantly decay into a pair of the heaviest kinematically accessible SM particles, resulting in final states containing metastable particles for the scalar mass $m_{S} > 2m_{\mu}$. Figure~\ref{fig:Neff-scalar} illustrates the impact of $S$ on $\Delta N_{\rm eff}$ as a function of $m_{S}$ and lifetime $\tau_{S}$. This and the following results are obtained using the unintegrated neutrino Boltzmann solver. We concentrate on the mass range $m_{S}\lesssim 2m_{\pi}$, where the scalar abundance is large enough to significantly affect the Early Universe's plasma~\cite{Fradette:2018hhl,Akita:2024ork}. There, it mainly decays into a pair of electrons or muons. The realistic setup, accounting for annihilation and nucleon interactions, shows a significant reduction in $|\Delta N_{\rm eff}|$ compared to the standard assumption of inevitable decay. Similarly to the toy model case, in the mass range $2m_{\mu} < m_{S} \lesssim 2m_{\pi}$, the sign of $\Delta N_{\rm eff}(\tau)$ changes due to the interplay between energy injection into neutrinos and the EM plasma.

\begin{figure}[t!]
    \centering
    \includegraphics[width=\linewidth]{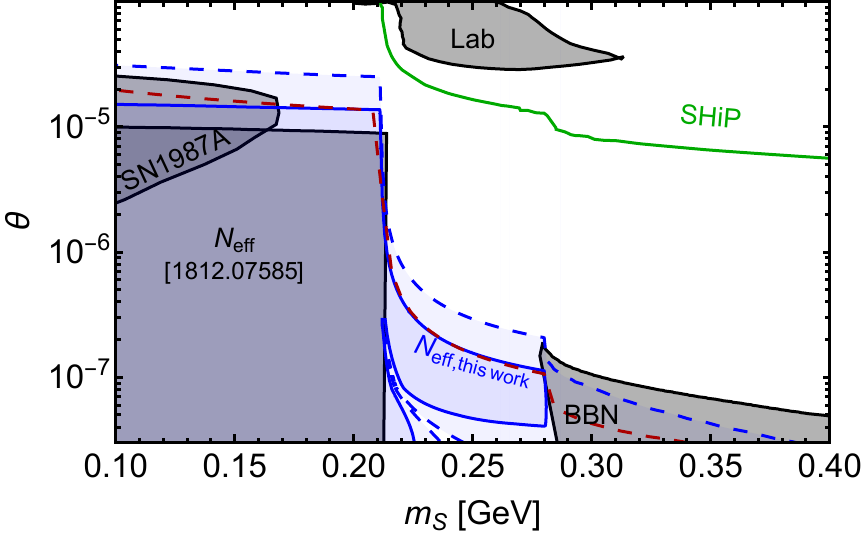}
    \caption{Parameter space of Higgs-like scalars in terms of the mass $m_{S}$ and the coupling to SM particles $\theta$. The plot shows various cosmological constraints (SN1987A, $N_{\rm eff}$, BBN) from ref.~\cite{Fradette:2018hhl}, laboratory bounds from ~\cite{Beacham:2019nyx}, and the sensitivity of the SHiP experiment from~\cite{Aberle:2839677}. The excluded region based on our $N_{\rm eff}$ calculation from CMB corresponds to the blue domain, whereas the dashed blue region envelopes the projected sensitivity of the measurements by the Simons observatory~\cite{SimonsObservatory:2018koc}. The dashed red line shows the iso-contour of the fixed lifetime $\tau_{S} = 0.1\text{ s}$.}
    \label{fig:constraints-scalar}
\end{figure}

The impact of scalars and other LLPs on $N_{\rm eff}$ can be used to constrain their couplings using CMB observations. In fig.~\ref{fig:constraints-scalar}, we show the parameter space in terms of mass and coupling to the Standard Model particles -- the mixing angle $\theta$. To compare it to Fig.~\ref{fig:Neff-scalar}, we need to relate the scalar lifetime to $\theta$: $\tau_{S} = \Gamma^{-1}_{S}$, where $\Gamma_{S} = f(m_{S})\cdot \theta^{2}$ is the decay width. $\Gamma_{S}$ rapidly increases at the di-muon and di-pion thresholds because corresponding decay modes become kinematically accessible. On the other hand, cosmological constraints are typically iso-contours of a fixed lifetime. Hence, they experience a sharp drop in $\theta$ at $m_{S} > 2m_{\mu}$.

We show laboratory constraints from~\cite{Beacham:2019nyx} and state-of-the-art cosmological constraints from~\cite{Fradette:2018hhl}. The latter includes the constraints on $N_{\rm eff}$, obtained using simplified analytic calculations. Our more complete analysis allows us to fill the gap in the mass domain $2m_{\mu}<m_{S}<2m_{\pi}$, which has been present in previous publications~\cite{Fradette:2018hhl}.

Our $N_{\rm eff}$ constraints are obtained by converting the results of fig.~\ref{fig:Neff-scalar} to the ($m_S,\theta$) plane and utilizing the 95\%CL Planck constraints $N_{\rm eff} = 2.99^{+0.33}_{-0.34}$. In addition, we show the potential of future CMB measurements by the Simons Observatory to explore broader parameter space. We improve the accuracy of the calculations of~\cite{Fradette:2018hhl} for the masses $m_{S} < 2m_{\mu}$ and extend them to higher masses to demonstrate that they compete with other probes, such as BBN.
\footnote{Our results would also affect the impact of LLPs on BBN. We leave a detailed investigation of this question for future work.}

Overall, with the other cosmological and astrophysical constraints, the CMB constraints complement the laboratory bounds, defining the target parameter space for future experiments such as the recently approved SHiP~\cite{Aberle:2839677}.

\bigskip

\textbf{Conclusion.} We have identified a crucial oversight in previous studies of the impact of Long-Lived Particles (LLPs) on cosmic neutrinos: the potential for metastable particles such as muons, pions, and kaons produced by decays of LLPs to annihilate or interact with nucleons before decaying. These effects can significantly alter the expected impact on primordial neutrinos, reducing $\Delta N_{\rm eff}$, mitigating spectral distortions, and inducing an asymmetry between the energy distributions of neutrinos and antineutrinos. Our findings necessitate a revision of cosmological studies on broad new physics models. With the analysis presented in this letter, we have studied in detail the effects on $\Delta N_{\rm eff}$ and spectral distortions; the neutrino-antineutrino energy asymmetry as a result of decays into charged kaons will be analyzed in future work.

To facilitate further research, we provide publicly accessible computational tools that incorporate the dynamics of $Y$ particles in a model-independent manner and integrate it into the neutrino Boltzmann solver.

\bigskip

\textbf{Acknowledgements.} This work has received support by the European Union’s Framework Programme for Research and Innovation Horizon 2020 under grant H2020-MSCA-ITN-2019/860881-HIDDeN, JSPS Grant-in-Aid for Scientific Research KAKENHI Grant No.~24KJ0060. K.A is grateful to Maksym Ovchynnikov and Thomas Schwetz for the hospitality during the stay at the Institute for Astroparticle Physics, KIT. The authors thank Miguel Escudero for carefully reading the manuscript and for providing useful comments.

\bibliography{bib.bib}

\end{document}